# Hydrogen trapping and diffusion in sub-stoichiometric vanadium and niobium carbide precipitates in high-strength steels

Xiaohan Bie[a], Baihua Ren[a], Xiao Zhou[a], Salim Brahimi[a], Stephen Yue[a], Jun Song[a,*]

[a] Department of Mining and Materials Engineering, McGill University, Montréal, Québec H3A OC5, Canada

## Abstract

High-strength steel is a structural metal crucial for load-bearing components yet is known to be highly susceptible to hydrogen embrittlement (HE). Vanadium (V) and niobium (Nb) containing precipitated carbides introduce strong hydrogen traps to immobilize hydrogen, thus mitigating HE. However, variations in intrinsic vacancy concentrations in these carbides affect hydrogen thermodynamics and kinetics but remain poorly understood. Employing first-principles calculations, hydrogen trapping and diffusion in V/Nb carbides were investigated. Hydrogen dissolution energies are composition-dependent, revealing a transition from reversible to irreversible trapping with increasing carbon vacancy content, prescribed by the strength of covalent bonds with neighboring V/Nb atoms. Meanwhile, the diffusion energy barrier decreases with increasing carbon vacancy content, attributed to changes in vacancy patterns within carbides. The findings contribute new and critical knowledge for understanding hydrogen trapping and diffusion in sub-stoichiometric V/Nb carbides, providing valuable guidance for process and composition innovation of high-strength alloy steels for better HE resistance.

## 1. Introduction

High-strength steels are critical structural materials for many industrial applications, particularly in load-bearing components, because they offer excellent strength, safety, and efficiency [1]. Furthermore, their excellent strength-to-weight ratio makes them highly desirable for achieving low-emission goals [2, 3]. In addition, high-strength steels are widely used as pipeline materials for the transportation of petroleum and natural gas [4, 5], and are considered an economical option for transporting and storing gaseous hydrogen in the emerging hydrogen economy [6]. However, the application of high-strength steels is greatly hindered by their susceptibility to hydrogen embrittlement (HE), a phenomenon in which the exposure to hydrogen significantly degrades the material's mechanical properties and/or induces premature material failure. Even at ultra-low hydrogen concentration, e.g., down to a few wt. ppm, high-strength steels may still suffer from HE to experience drastic mechanical degradation [7, 8]. Furthermore, high-strength steels often become more susceptible to HE as the strength increases, which discourage the effort to further innovate high-strength steels with higher strength.

HE is further manifested by the common existence of hydrogen in service environments and hydrogen being a typical product from corrosion [9, 10]. Since its initial discovery [11], HE has been an active research topic in both academia and industry [12-14]. Over the years, several competing models, based on different notions of how hydrogen interacts with metal lattice and microstructure to affect the cohesion strength and/or dislocation behaviors, have been proposed to explain HE in metals, with some notable ones being hydrogen-enhanced decohesion (HEDE) [15, 16], hydrogen-enhanced local plasticity (HELP) [17, 18], hydride formation and cleavage [19, 20]. Though these mechanisms, individually each provide useful mechanistic insights towards understanding of HE, there are conflicts among their interpretations of HE, and significant discrepancy and even contradiction exist in experimental observations [21-25]. There have also been many studies suggesting that different HE mechanisms may prevail under different conditions or even interplay with each other to operate simultaneously [26-29], yet no clear consensus exists. To-date, there remain no reliable quantitative prediction of HE, in its occurrence or degree of damage, even for simple metallic systems.

The puzzle around HE is more entangled in engineering materials like high-strength steels, where there is rich variation in alloy composition, heat treatment and complex microstructures therein. Nonetheless, based on the experimental results, empirically it is generally believed that the occurrence of HE in high-strength steels is mainly attributed to hydrogen dissolved in bulk lattice or at reversable trap sites, i.e., interstitial sites in microstructural entities with low and medium trapping energies [30]. These hydrogen atoms are termed as mobile or diffusible hydrogen as during loading they are able to diffuse under ambient conditions to strained regions, to accumulate into sufficient quantity to modify plasticity and/or decohesion behaviors of the material to induce embrittlement. Revolving around such notion, in engineering practices, methods developed to mitigate HE damage in high-strength steels often attempt to either slow down hydrogen diffusion [31] or engineer more strong traps to immobilize and reduce diffusible hydrogen [32]. On the front of engineering strong hydrogen traps, precipitates play an important role. In particular, metal carbides are widely utilized as effective agents for such purpose besides their normal usage as a means for strength enhancement [33-39]. Among various metal carbides, vanadium (V) and niobium (Nb) containing carbides are non-stoichiometric and capable of accommodating a large number of carbon vacancies [32, 40-42], which are strong hydrogen traps in steels [32, 41, 43-46]. Studies have shown that vacancy distribution and composition in precipitated carbides in steels largely depend on both tempering temperature and carbon activity [32, 36], which consequently would affect hydrogen trapping characteristics of carbides. For instance, hydrogen trapping capacity of V or Nb containing steel was found to increase significantly after carbide precipitation was induced by tempering [47, 48]. Furthermore, Takahashi et al. reported that the transformation of precipitated VC to $V_4C_3$ led to a significant increase in both hydrogen trapping energy and capacity, indicating that the precipitated carbide's sub-stoichiometry plays a key role in hydrogen trapping [32]. Meanwhile, recent studies have also indicated that the kinetics of hydrogen in carbides are also strongly composition dependent. Several atom probe tomography (APT) experiments were conducted to identify trapped hydrogen at carbide precipitates, revealing that an increased number of carbon vacancies enhances the hydrogen trapping capacity while simultaneously reducing the hydrogen diffusion barriers within carbides [32, 49-51].

However, despite experiments indicating that hydrogen trapping and diffusion behaviors at carbide precipitates are composition (and carbon vacancies) dependent, such dependence has not been quantitatively analyzed in connection to sub-stoichiometry in carbide compounds, nor are there dedicated studies clarifying the fundamental origins underlying the dependence.

The present study aims to address the knowledge gap by directly investigating hydrogen trapping and diffusion characteristics in different sub-stoichiometries carbides formed in the presence of carbon vacancies. Focusing on V-/Nb-containing carbides as the representing carbide systems, hydrogen trapping energies and diffusion barriers in these carbide compounds were systematically studied, employing first-principles calculations. The stable interstitial sites for hydrogen in ground-state sub-stoichiometric V/Nb carbides were identified to coincide with the carbon vacancy centers, which were shown to serve as effective hydrogen trap sites. The hydrogen dissolution energies at these sites were calculated, revealing dependence on carbide composition and a transition from reversible to irreversible trapping as the carbon vacancy content increases. The electronic and structural origins underlying the variation in hydrogen dissolution energy were then clarified. In addition to the energetics of hydrogen, the diffusion pathways and energy barriers of hydrogen, in those sub-stoichiometric carbides were unveiled, also found to be critically dependent on the carbide composition. The present study contributes critical knowledge for hydrogen trapping and diffusion within sub-stoichiometric V/Nb carbides, of significant value to engineering of alloy carbide precipitates in high-strength steels for better HE resistance.

# 2 Computational Methodology

## 2.1 First-principles calculations

We conducted first-principles calculations using the Vienna Ab initio Simulation Package (VASP) [52, 53] to determine the dissolution energies of hydrogen in different Nb-/V-containing carbide systems. In our previous study [54], the stable sub-stoichiometric compounds as the amount of carbon vacancies varies were predicted, based on which the simulation supercells of carbide compounds were constructed, listed in **Table SI** in Supplementary Material. For simplicity, below we denote the carbide as $M_mC_n$ where $M$ = Nb or V while $m$ and $n$ indicate the counts of M atoms and C atoms within the compound respectively. Kohn-Sham Density-functional theory (DFT) was utilized to compute the total energies of different systems [55, 56]. As for potentials, the electronic correlation and exchange effects are studied using generalized gradient approximation (GGA) of PW91 functional [57]. The hydrogen dissolution energy was computed as the energy difference between a carbide supercell with one hydrogen atom and the corresponding supercell without it. For all the carbide systems, the supercell size was tested to ensure it is sufficient to prevent hydrogen-hydrogen image interactions so that there is no size dependence in the hydrogen dissolution energy. The plane-wave cut-off energy was set to 500 eV, and the convergence criteria for energy and force were set to be $10^{-8}$ eV per simulation cell and 0.01 eV/Å respectively, in all calculations. Gamma centered $k$-point meshes are utilized for sampling Brillouin Zone, with convergence tests for $k$-point meshes performed with a convergence criterion of 1 mV/atom. The detailed k-meshes are listed in **Table SII** in Supplementary Material. Moreover, it is worth noting that in the calculations described above, the zero-point energy (ZPE) [58] correction was not considered. Nevertheless, given that for light atoms such as hydrogen, ZPE could influence their dissolution energies [58], preliminary calculations incorporating the ZPE correction have been carried out, with the results indicated negligible influence of ZPE (see additional details in Supplementary Material)

## 2.2 Bonding characterization

To better understand hydrogen trapping in the carbide compound, we analyzed the characteristics of relevant bonding involved. The nature of bonding between two atoms can be discerned by examining the disparity in their electronegativities. When the electronegativity difference falls below a benchmark value of about 1.7, the bond may be classified as predominantly covalent [59].

In a more quantitative fashion, the electronegativity difference (denoted as $\Delta X$) may be fed into the Pauling equation, to calculate the percentages of ionic and covalent bonding (denoted as $P_{ion}$ and $P_{cov}$ respectively) [60],

$$P_{ion} = 1 - \exp[-0.25 \times (\Delta X)^2], \quad P_{cov} = 1 - P_{ion} \tag{1}$$

For the carbide system, $\Delta X = X_H - X_M$, where $X_H$ and $X_M$ respectively denotes the electronegativities of the hydrogen and metal (i.e., V or Nb) atoms.

Moreover, further bonding characterization was performed by projecting plane waves to local orbital basis functions for extracting crystal orbital Hamilton population (COHP) [61] to dissect the band structure energy into orbital-based pairwise interactions. The computer program LOBSTER (Local Orbital Basis Suite Towards Electronic-Structure Reconstruction) was selected for doing the projection [62, 63]. In the COHP method, the band structure energy is partitioned into a series of pairwise orbital contributions [61]. The bonding, antibonding and nonbonding interactions between two selected atoms can be therefore revealed. The net bonding characteristics can be received when COHP values are integrated up to Fermi level ($E_f$) [64]

Further to COHP, the projected COHP, namely pCOHP was also calculated [62]. As an extension to the original COHP, pCOHP allows the analysis of chemical bonds based on plane-wave DFT calculations [62]. It operates by projecting the delocalized plane wave functions onto a local atomic basis set [62]. This projection allows the reconstruction of the Hamiltonian and density matrices in a localized representation, thus enabling COHP-like bond analysis in the plane-wave formalism.

**2.3 Nudged elastic band (NEB) calculations**

The NEB method [65] was employed to investigate the local hydrogen diffusion pathways and associated energy barriers associated in carbides. In such NEB calculation, images were generated with hydrogen interpolated between the two interstitial sites corresponding to the start and final states of hydrogen migration. The distance between neighboring hydrogen atoms in the interpolated images was kept under 1Å to ensure convergence of the results. The minimum energy path (MEP) connecting the initial and final states was determined using the harmonic approximation to transition state theory [66, 67]. To ensure the accurate identification of saddle points along the hydrogen migration path, the climbing image NEB (CI-NEB) was employed [68].

## 3 Results and discussions

### 3.1 Hydrogen trapping energetics in carbides

Hydrogen adsorption in a perfect, vacancy-free V/Nb carbide has been previously examined [69, 70], where hydrogen was found to reside at the interstitial sites. There are three types of interstitial sites in vacancy-free V/Nb carbides, namely the center of the V/Nb triangle, referred to as the triangle center (abbreviated as TC), body-centered (BC) interstitial and face-centered (FC) interstitial, as depicted in **Figure 1**. Among these three sites, the preferred one for hydrogen adsorption is the TC site as displayed in **Figure 1(c)** [69].

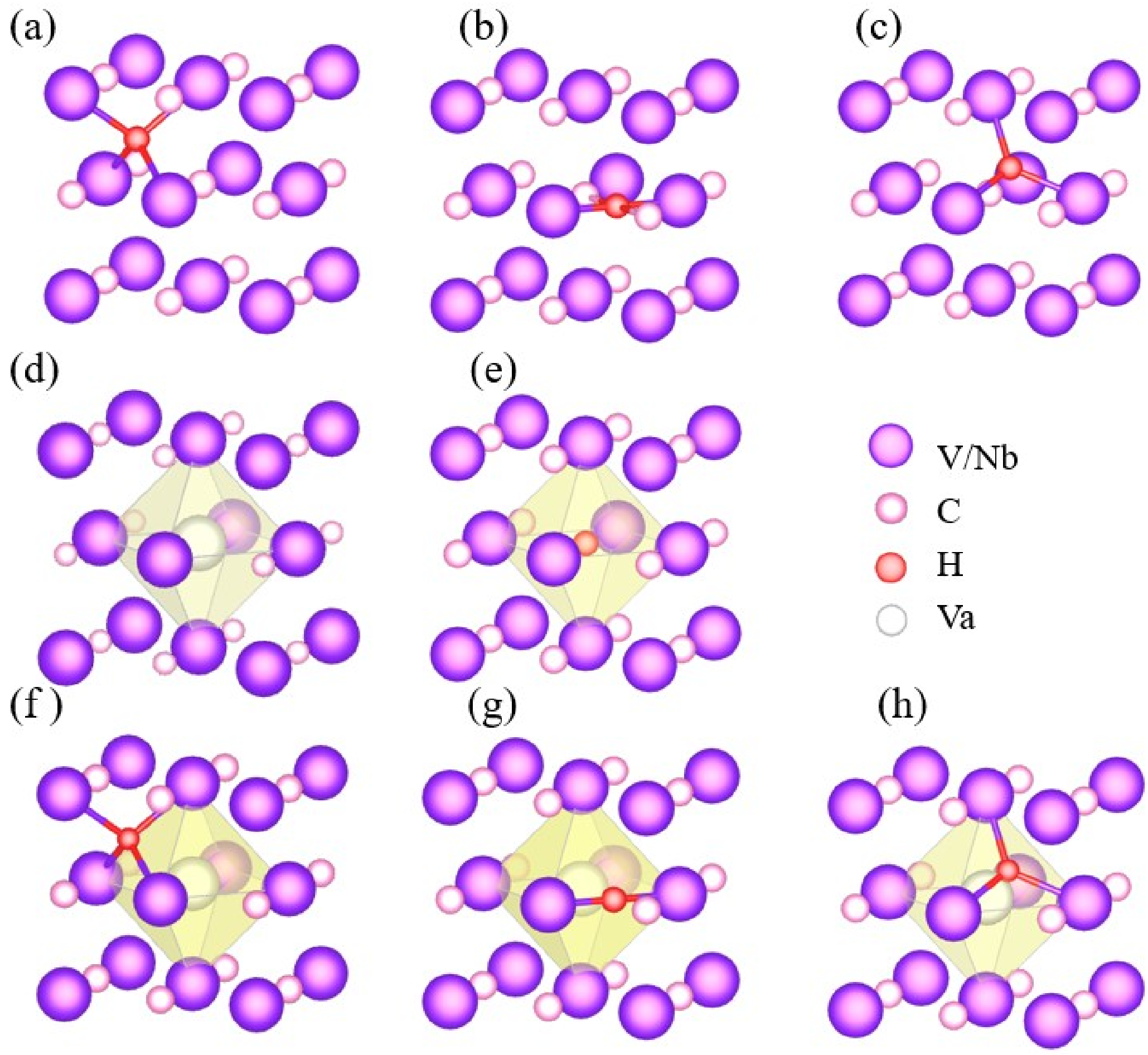

**Figure 1.** Three distinct interstitial sites for hydrogen adsorption in pristine defect-free VC/NbC, being the (a) body center (BC), (b) face center (FC), and (c) triangle center (TC) sites. (d) and (e) show the octahedron unit enclosing an isolated vacancy in VC/NbC, and the interstitial site located at the center of such octahedron unit ($VO_c$) respectively, while (f)-(h) show other possible interstitial sites within the close vicinity of the vacancy, being body center site (Va-BC), face center site (Va-FC) and triangle center (Va-TC) respectively. The V/Nb, C and hydrogen atoms are colored purple, pink and red respectively, while the vacancy site (Va) is indicated by a gray circle.

The presence of carbon vacancies in the carbides induces new interstitial sites for hydrogen adsorption. Each carbon vacancy in the V/Nb carbides is surrounded by six metal atoms, forming an octahedron (as indicated in **Figure 1(d)**). Focusing on the simple case of a single carbon vacancy, the stable trapping site for a hydrogen atom is at the center of the octahedron ($VO_c$), which also coincides with the vacancy center, as illustrated in **Figure 1(e)**. It is also worth noting that we have also examined other possible interstitial sites within the close vicinity of the vacancy, including the body center (Va-BC), face center (Va-FC) and surface triangle center of the octahedron (Va-TC) near the vacancy, as shown in **Figure 1(f-h)**. However, these interstitial sites, other than the vacancy center ($VO_c$), all exhibit instability for hydrogen, and hydrogen atoms initially placed at these sites would just move to the $VO_c$ post relaxation. It is also worth mentioning that the above preference and stability of hydrogen at the vacancy center remain as the vacancy content increases in V/Nb carbides.

With the stable interstitial sites for hydrogen atoms identified, we then examined hydrogen trapping at the interstitial sites by evaluating the hydrogen dissolution energy, calculated as the following:

$$E^d = E(MC_xH) - E(MC_x) - \frac{1}{2}E(H_2) \tag{2}$$

where $E(MC_xH)$ is the total energy of a carbide compound with one hydrogen atom at the interstitial site, while $E(MC_x)$ represents the total energy of the reference, hydrogen-free carbide compound. $E(H_2)$ is the energy of a hydrogen gas molecule, whose value was determined to be -6.79 eV.

**Table 1.** Interstitial sites for hydrogen within the pristine vacancy-free VC/NbC, and those within close proximity to an isolated single carbon vacancy in VC/NbC. The corresponding values of hydrogen dissolution energy ($E^d$) are listed, in comparison to available data from the literature. Note that the hydrogen dissolution energy is 0.3 eV in the tetrahedral sites of pristine bcc Fe [71], and that no hydrogen dissolution values are listed for Va-BC, Va-FC and Va-TC sites because hydrogen is unstable at those sites.

| Hydrogen trapping sites | VC | | NbC | |
|---|---|---|---|---|
| | Hydrogen dissolution energy (eV) | | | |
| | This Study | Literature* | This study | Literature* |

| | | | | | |
|---|---|---|---|---|---|
| Pristine | BC | 2.06 | 2.111 [69], 2.078 [70] | 1.92 | 2.137 [70] |
| | FC | 1.82 | 1.812[69], 1.929 [70] | 1.88 | 2.236 [70] |
| | TC | 1.49 | 1.621[69], | 1.19 | |
| Single vacancy | $VO_c$ | -0.0481 | -0.074 [72], −0.07 [73] | 0.0128 | 0.02 [73] |
| | Va-BC | -- | | -- | |
| | Va-FC | -- | | -- | |
| | Va-TC | -- | | -- | |

* Necessary conversion was made as different formulae were used in literature studies.

Starting with pristine vacancy-free V/Nb carbides, the hydrogen dissolution energies at the different trapping sites were obtained and shown in **Table 1**. The results obtained by us are in good agreement with the previously reported values, and it is apparent that hydrogen adsorption in vacancy-free V/Nb carbides is much energetically unfavorable compared to the pristine bcc Fe. On the other hand, carbon vacancies are found to introduce stable trap sites for hydrogen. For an isolated vacancy in the V/Nb carbides, the hydrogen dissolution energy at the vacancy center ($VO_c$) assumes values of -0.0481 eV and 0.0128 eV for VC and NbC system respectively, much lower than that for hydrogen at a tetrahedral site in the pristine bcc iron lattice (0.30 eV).

Moreover, for hydrogen trapping at vacancies, the neighboring interstitial sites become unstable, regardless of the carbon vacancy content, and the hydrogen dissolution energy at $VO_c$ sites does vary as the vacancy content changes. **Figure 2** shows the hydrogen dissolution energy at the $VO_c$ site in the various sub-stoichiometric V/Nb carbide compounds predicted in our previous study [54]. To put our discussion in the direct context quantitatively linked to hydrogen trapping, we take the reference hydrogen dissolution energy (i.e., hydrogen at a tetrahedral site in the pristine bcc iron lattice), denoted as $E_0^d$ (=0.30 eV, indicated in **Figure 2** as the pink dashed line) as the reference [71], and define the hydrogen trapping energy $E_t = E^d - E_0^d$. It is important to note that although the dissolution energy $E^d$ and hydrogen trapping energy $E_t$ are closely related. They have distinct physical meaning and are defined with respect to different reference states, being the energy states of hydrogen in a hydrogen molecule and pristine bcc iron lattice respectively. The trapping energy $E_t$ provides a direct metric for quantitative comparison of trapping efficacy among different trapping sites in steel.

Hydrogen trapping sites (abbreviated as traps thereafter) with $E_t$ less than 0.31 eV (30 kJ/mol) are typically considered to be weak trapping [74], and the corresponding dissolution energy boundary is denoted as $E_{WTB}^d$, depicted as the blue dash dotted line in **Figure 2**. On the other hand, hydrogen traps with $E_t$ greater than 0.62 eV (60 kJ/mol) [73-75] are considered to be irreversible trapping, and its corresponding dissolution energy boundary is denoted as $E_{STB}^d$, indicated by the purple dotted line in **Figure 2**. It is important to note that the reversibility/irreversibility of hydrogen trapping is dependent on the actual service conditions. In fact, strictly speaking, one may not define a single energy cut-off value to separate reversible or irreversible traps. The afore-mentioned dissolution boundary energy values were employed more for convenience of classifying hydrogen traps into different categories. They are intrinsically empirical in nature. In particular, the energy values we used are based on the classification of traps in steel by Pressouyre [7]. From the results, we see that clearly the hydrogen dissolution energy decreases with increasing vacancy concentration in carbides, starting initially close to $E_{WTB}^d$, but dropped below $E_{STB}^d$ when the carbon vacancy concentration is beyond 25% in carbon sublattice (i.e., for $M_4C_3$ and $M_3C_2$ carbides). As vacancy concentrations increase, there is a simultaneous elevation not only in hydrogen trapping energies but also in the overall quantity of entrapped hydrogen atoms, attributable to the increased presence of vacancies. This aligns with experimental findings indicating a heightened strength of hydrogen traps when the precipitated V carbides undergo a transformation from VC to $V_4C_3$, a transformation that was shown to result in an increase in trapping energy from 24.8 ± 5 kJ/mol to 59.6 ± 10 kJ/mol [32]. Beyond the augmentation in trap strength, the trapping capacity also experiences a two- to fourfold amplification following the carbide transformation [32].

The results provide critical information on the characteristics of V/Nb carbides as hydrogen trapping agents, demonstrating that these carbide systems would only act as strong traps when there is sufficient carbon vacancy concentration therein (i.e., ≥ 25% in VC/NbC's carbon sublattice), while otherwise would provide moderate or weak traps. As typically the strong traps are expected to be more beneficial for HE mitigation, $M_4C_3$ and $M_3C_2$ carbides are thus more desirable in reducing the HE susceptibility as they could enable a more substantial irreversible entrapment of hydrogen atoms. Another interesting observation is that in a given ground-state sub-stoichiometric carbide, all carbon vacancies within the carbide share the same chemical environment [77]. Therefore, the dissolution energy of hydrogen in $VO_c$ sites within one carbide is the same, as illustrated by a single $E^d$ value for each carbide shown in **Figure 2**. This is because that the $VO_c$ sites in sub-stoichiometric carbides are intrinsic structural units. Consequently, these $VO_c$ sites do not possess the randomness like the conventional carbon vacancies induced by, e.g., by thermal activation or deformation. Of course, in actual carbide compounds, variation in the local chemical environments surrounding the $VO_c$ sites can happen, which would in turn render statistical fluctuation in the $E^d$ value measured.

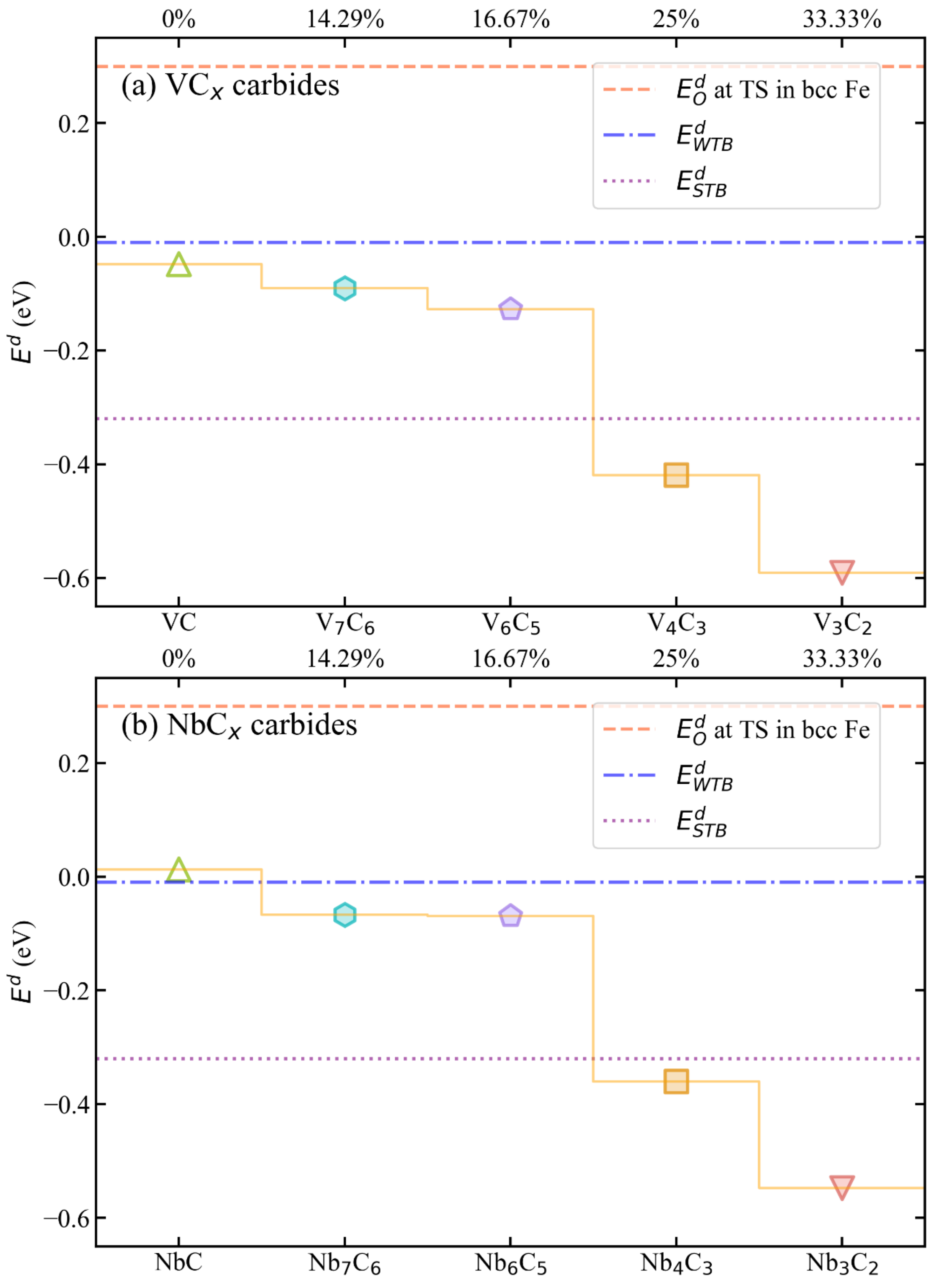

**Figure 2.** $E^d$ value for hydrogen at the $VO_c$ site in different (a) V carbide and (b) Nb carbide compounds with varying carbon vacancy concentration. For VC and NbC, $E^d$ signifies hydrogen trapping at a single vacancy (open triangle symbol). The pink dashed line, blue dash-dotted line and purple dotted line indicate hydrogen dissolution energies $E_0^d$ in bcc Fe, the weak trapping boundary $E_{WTB}^d$, and the strong trapping boundary $E_{STB}^d$, respectively.

**Table 2**. Hydrogen dissolution $E^d$ and activation energy $E^a$ values for hydrogen adsorption at carbon vacancies and diffusion between adjacent carbon vacancies in different carbide compounds. ED_Vac denotes the electron density value at the carbon vacancy center. $E^{ela}$ and -IpCOHP signify the elastic deformation energy and bonding energy components of hydrogen at the vacancy center. Note that the standard deviations for these quantities are negligible, thus not provided here.

| System | $E^d$ (eV) | $E^{ela}$ (eV) | ED_Vac (eV/Å$^3$) | -IpCOHP (eV) | $E^a$ (eV) |
|---|---|---|---|---|---|
| $V_3C_2$ | -0.591 | 0.011 | 0.123 | 0.83 | 1.05 |
| $V_4C_3$ | -0.420 | 0.018 | 0.109 | 0.82 | 0.80 |
| $V_6C_5$ | -0.127 | 0.022 | 0.091 | 0.80 | 2.20 |
| $V_7C_6$ | -0.090 | 0.025 | 0.087 | 0.78 | 2.43 |
| VC* | -0.048 | 0.027 | 0.089 | 0.80 | – |
| $Nb_3C_2$ | -0.548 | 0.006 | 0.119 | 0.95 | 0.74 |
| $Nb_4C_3$ | -0.360 | 0.013 | 0.110 | 0.92 | 0.55 |
| $Nb_6C_5$ | -0.069 | 0.019 | 0.098 | 0.87 | 2.31 |
| $Nb_7C_6$ | -0.067 | 0.010 | 0.096 | 0.86 | 2.33 |
| NbC* | 0.0128 | 0.128 | 0.0945 | 0.88 | – |

* Denotes a single vacancy in VC/NbC

## 3.2 Factors affecting characteristics of hydrogen trapping

### 3.2.1 Electron density and charge in vacancy prior to hydrogen adsorption

Many previous studies have suggested that disparity in hydrogen dissolution energies can often be linked to difference in the electronic states at hydrogen adsorption sites [78]. In this regard, we analyzed the associated electronic structures and bonding characteristics of hydrogen at traps (i.e., $VO_c$) across different carbide compounds. In particular, we determined the electron density, referred to as $n_e$ hereafter, at a $VO_c$ site prior to hydrogen insertion (with detailed electron density data listed in **Table 2**).

Examining these electron density data, we found that they showed a notable correlation with the dissolution energies of hydrogen (see **Figure S1** in **Supplementary Material**). Motivated by this, we computed the Voronoi volumes (denoted as $\Omega_V$) of vacancies prior to hydrogen insertion. The product of $\Omega_V$ and the afore-computed electron density $n_e$, i.e., $n_e\Omega_V$, provides an effective means to approximate the electron charge associated with a trap at $VO_c$. Using $n_e\Omega_V$ as a descriptor to capture the variation in electron charge at $VO_c$ as the stoichiometry changes, we plot the hydrogen dissolution energy $E^d$ versus $n_e\Omega_V$ in **Figure 3**. The plot reveals a clear linear relationship, which may be mathematically depicted as:

$$E^d = k_0 n_e \Omega_V + b \tag{3}$$

where $k_0$ and $b$ are constants, whose values were obtained via least-square fitting and listed in **Table 3.**

**Table 3.** The values of parameters $k_0$, $b$, and $\Omega_V$ (c.f. **Eq. 3**), number of electrons for V/Nb ($N_e$), $a$ and c (c.f. **Eq. 4**), obtained from least-squares fitting for the carbide systems.

| System / Parameter | $k_0$ (eV/electron) | $b$ (eV) | $\Omega_V$ (Å$^3$) | $N_e$ | $a$ | c (eV) |
|---|---|---|---|---|---|---|
| $VC_x$ | -1.66 | 1.23 | 9.0 | 23 | -11.72 | 9.16 |
| $NbC_x$ | -1.98 | 2.13 | 11.3 | 41 | -6.00 | 5.14 |

From **Figure 3**, it can be noted that $n_e\Omega_V$ increases as the carbon vacancy concentration increase, and in the meantime, increasing $n_e\Omega_V$ results in decreasing hydrogen dissolution energy, thus stronger hydrogen trapping at the vacancy center. These observations underly the critical role of electron charge to hydrogen trapping behaviors at vacancies and varying trapping strength in different carbide compounds. In other words, the charge associated with a vacancy contributes significantly to hydrogen trapping therein, consistent with existing observations [78].

Meanwhile, we can note from **Table 3** that V and Nb carbide systems exhibit different $k_0$ values. The Nb carbide system exhibits a more negative $k_0$, indicating that increasing electron charge leads to a more significant stabilization of hydrogen compared to the V carbide system. This might be attributed to the nature of the $d$-orbitals and hybridization. Nb has a larger $d$-orbital (4*d*) than V (3*d*), and their difference in the spatial extents and energies of $d$-orbitals could lead to different degrees of orbital overlap and hybridization with the 1$s$ orbital of hydrogen. As explained later (in **Section 3.2.2**), the correlation between the $d$-band center and hydrogen dissolution energy provides electronic structure level evidence supporting this argument.

The linear relationship shown in **Eq. 3**, though empirical in nature, is of practical significance and enables a robust assessment of hydrogen binding strength using $n_e$ and $\Omega_V$, two quantities readily accessible from calculations. The relationship is also in concert with the fact that at carbon vacancies, the local electron density is typically below the optimal value preferred by hydrogen, and increasing it thus promotes hydrogen binding at therein [79]. Our findings also align with previous reports of hydrogen preferentially occupying interstitial sites characterized by high pre-existing charge densities in metals such as V, Nb, Fe, and Mo [78].

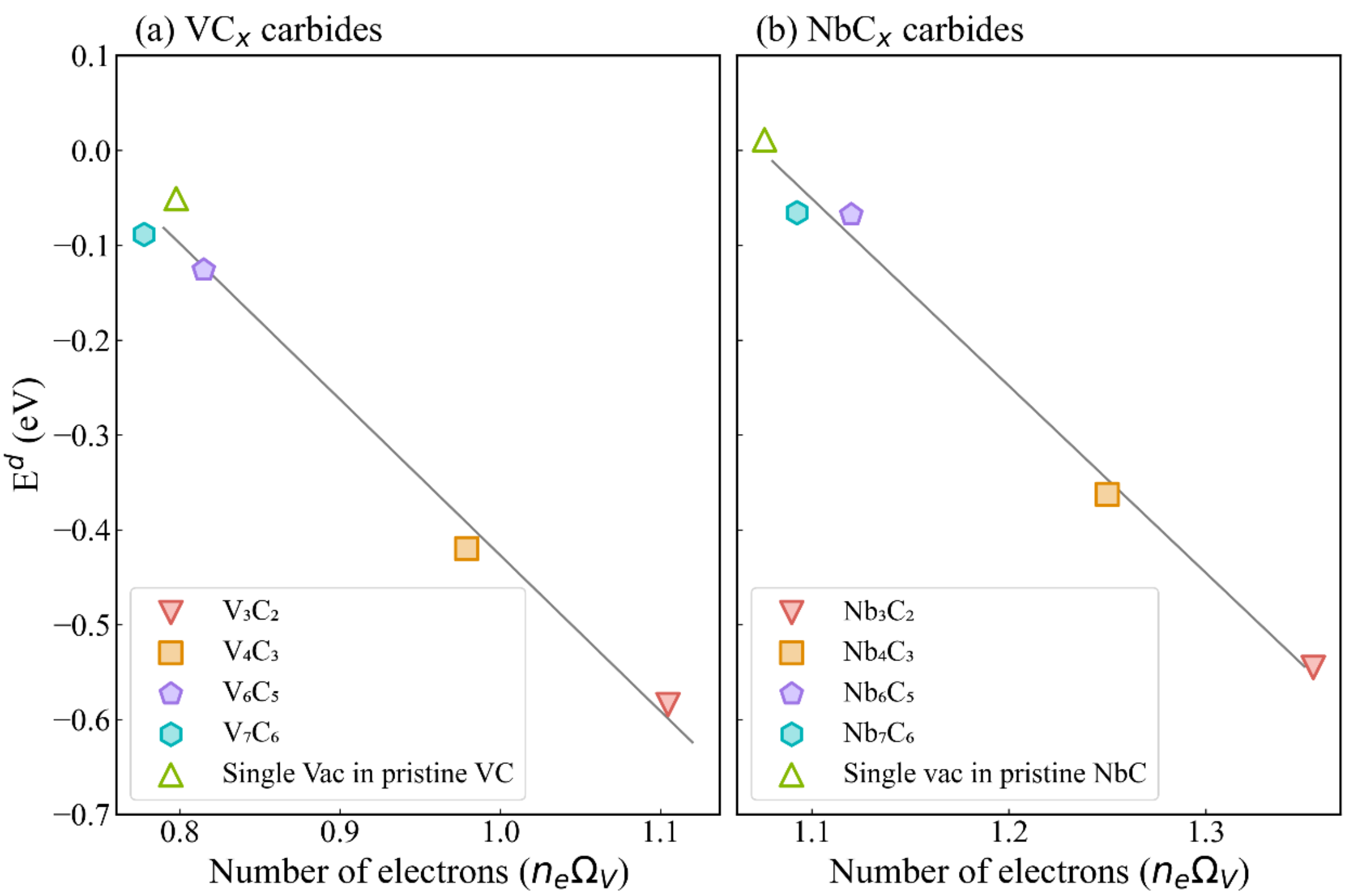

**Figure 3.** Plots of the hydrogen dissolution energy ($E^d$) of hydrogen trapped at the vacancy versus the term $n_e\Omega_V$, representing the electron charge associated with the interstitial site at the vacancy, for (a) VC$_x$ and (b) NbC$_x$ carbide systems, revealing a linear relationship, indicated by the black line. Note that the green open symbols indicate a single vacancy in the pristine MC (M = V or Nb).

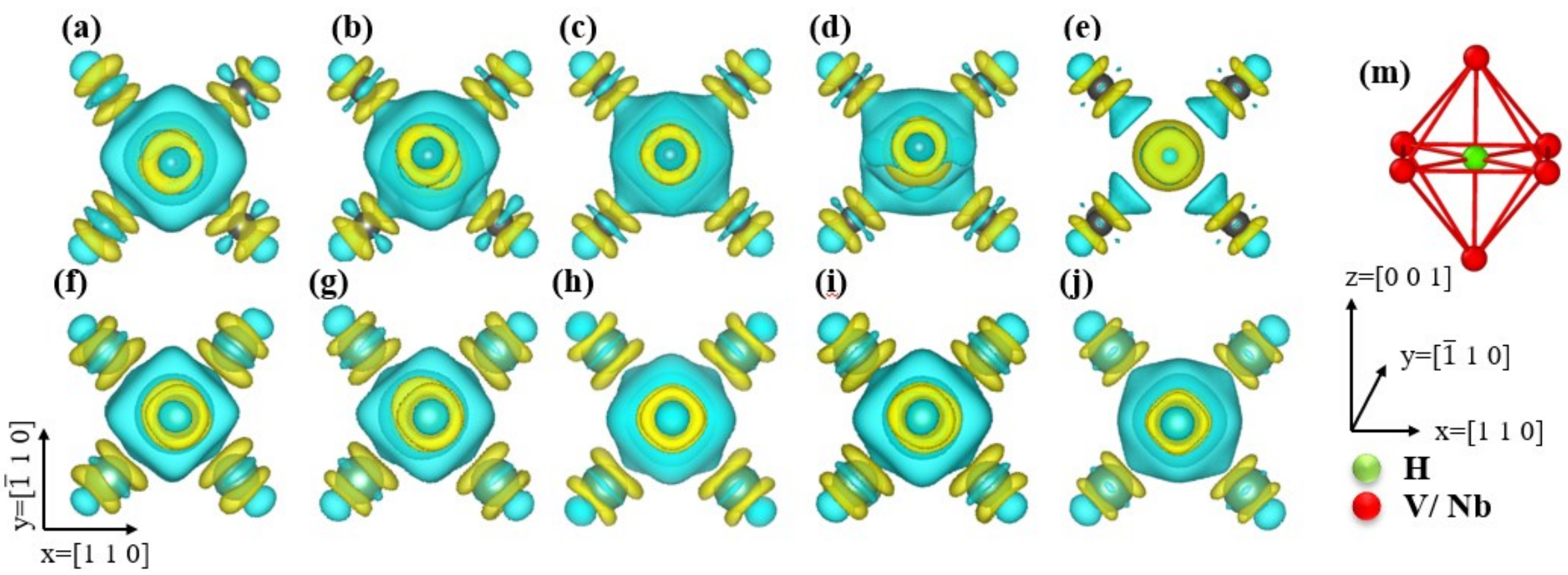

**Figure 4.** Electron density difference iso-surfaces after hydrogen insertion in carbon vacancy in (a) $V_3C_2$, (b) $V_4C_3$, (c) $V_6C_5$, (d) $V_7C_6$, (e) single vacancy in pristine VC, and (f) $Nb_3C_2$, (g) $Nb_4C_3$, (h) $Nb_6C_5$, (i) $Nb_7C_6$, (j) single vacancy in pristine NbC. (m) shows the octahedra of V/Nb surrounding the inserted hydrogen atom. The regions depicted in yellow represents electron gain, while those in blue denotes electron loss.

### 3.2.2 Charge transfer and electronic structure post hydrogen adsorption

In addition to the previous analysis of electron density and charge prior to hydrogen adsorption, we have also performed analysis of local bonding between hydrogen and V/Nb post hydrogen insertion. In the cases of Nb–hydrogen and V–hydrogen bonds, the electronegativity values of hydrogen, V and Nb are 2.20, 1.63 and 1.60 respectively [80], the electronegativity difference values for the hydrogen-Nb and hydrogen-V bonds are 0.57 and 0.60, which falls well below the conventional threshold of 1.7 [59] that delineates ionic bonding from covalent bonding. According to **Eq. 1**, the percentages of ionic character ($P_{ion}$ ) are calculated to be 0.078 and 0.086, respectively, while the corresponding covalent character ( $P_{cov}$) values are $0.922$ and 0.914, respectively, for hydrogen-Nb and hydrogen-V bonds, indicative of both being predominantly covalent [59, 80].

With the hydrogen dissolution energy demonstrated to strongly correlated with the local charge at vacancy (**Figure 3**), and charge transfer known to play a pivotal role in the formation and stability of chemical bonds between atoms [81], we examined the electron density difference induced by hydrogen insertion. Plots in **Figure 4** illustrate the electron density difference resulted from hydrogen insertion at a representative trap site, for different carbide compounds. For easy visualization, projected views of the electron difference, onto the (001) plane, are shown in the figure. The results showed significant charge transfer between the hydrogen atom and its neighboring metal atoms and that the charge transfer was predominantly localized around the metal octahedron enclosing the hydrogen atom. The charge transfer is a direct consequence of valence charge migration, wherein electrons shift from the metal atoms to the hydrogen atom, and can be attributed to the higher electronegativity of hydrogen compared to V/Nb atoms. Further to the charge transfer analysis shown in **Figure 4**, the associated projected density-of-states (PDOS) curves were also computed, with the results presented in **Figure 5.** Analysis of the PDOS results reveals that bonding between the inserted hydrogen and neighboring metal atoms primarily originates from the *s*-*d* orbital hybridization, between the 1$s$ orbital of hydrogen and the 3$d$ (V) or 4$d$ (Nb) orbital of the host atom. The peak at −7 eV below the Fermi level indicates significant orbital overlap, contributing to the formation of covalent bonding [64]. This observation aligns with previous studies that identified deep-level hybridized states between hydrogen 1$s$ and V 3$d$/Nb 4$d$ orbitals, which play a key role in stabilizing hydrogen in metals, as confirmed by hard X-ray photoelectron spectroscopy (HAXPES) [82, 83].

To further validate the hypothesis that *d* orbitals of V or Nb play a critical role in governing the hydrogen dissolution energy, we calculated the correlation between hydrogen dissolution energy and the *d*-band center of the metal atoms in different carbide systems [84]. The *d*-band center, denoted as $\varepsilon_d$, is evaluated as:

$$\varepsilon_d = \frac{\sum_i E_i \cdot N_d(E_i)}{\sum_i N_d(E_i)} \tag{4}$$

where $E_i$ are the discrete energy points relative to the Fermi level ($E_F$), and $N_d(E_i)$ is the PDOS of the *d* orbitals at $E_i$. **Figure 6** plots the relationship between the calculated metal *d*-band center, $\varepsilon_d$, and the hydrogen dissolution energy at the carbon vacancy site, clearly showing a linear relationship. Specifically, as the *d*-band center approaches the Fermi level, the hydrogen dissolution energy decreases, indicating stronger hydrogen trapping. This relationship follows the *d*-band theory proposed by Hammer and Nørskov [85, 86], describing an approximately linear correlation between the adsorption energy of adsorbates on transition metal surfaces and the *d*-band center of the metal. Although the *d*-band model was originally formulated for surface adsorption, its underlying principle of the interaction between adsorbate states and metal *d* states has been shown to also apply to hydrogen dissolution in metals [87]. Notably, even though our system involves carbides, the *d*-band model remains largely valid because the metal atoms predominantly govern the bonding interactions with hydrogen.

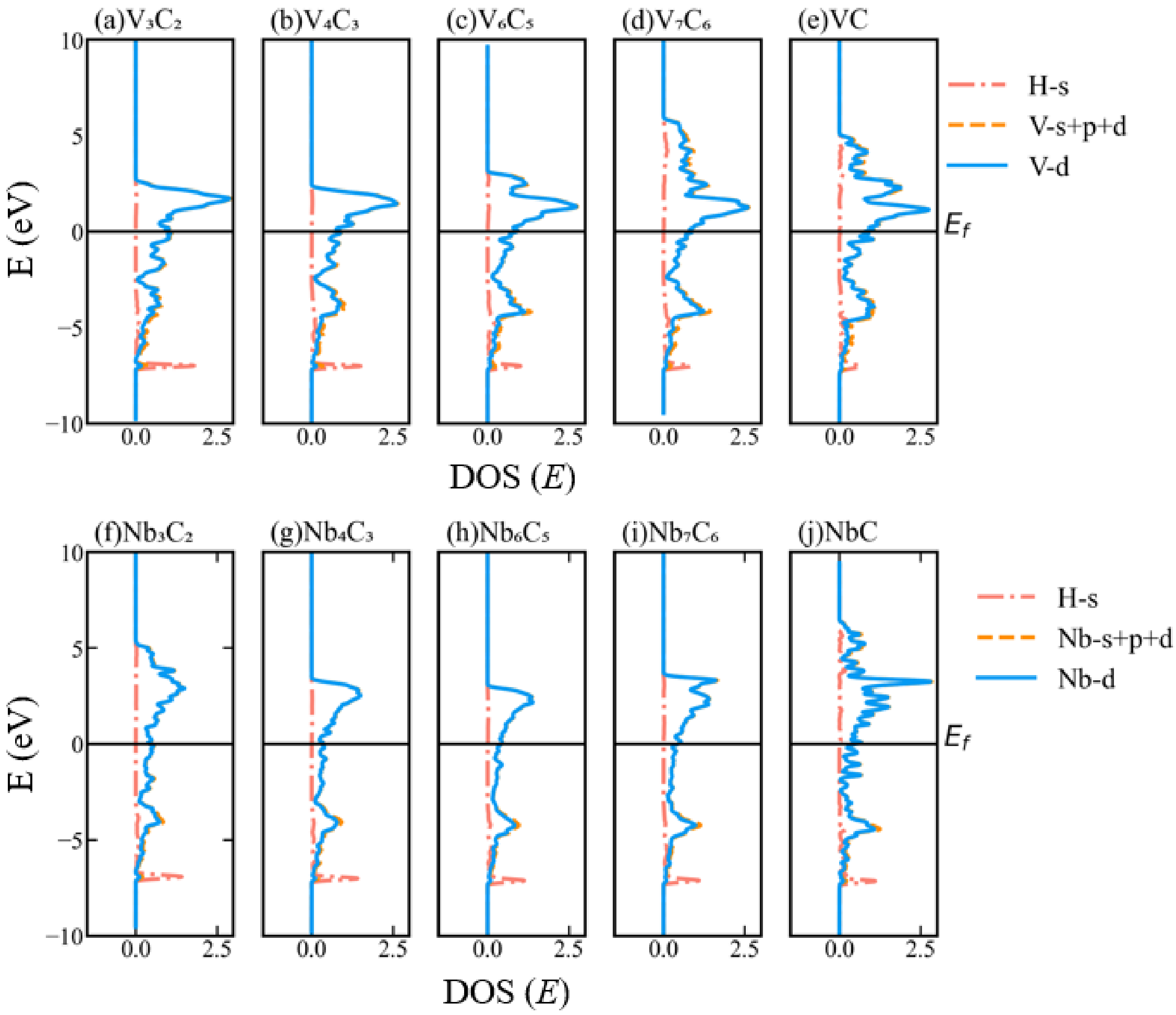

**Figure 5.** The density-of-states (DOS) curves for the adjacent hydrogen-V/Nb atoms within the metal octahedron (see **Figure 4(m)** for schematic illustration of the octahedron), for different $VC_x$/$NbC_x$ carbides. Note that (e) and (j) represent the cases of hydrogen occuping a single vacancy in pristine VC and NbC respectively.

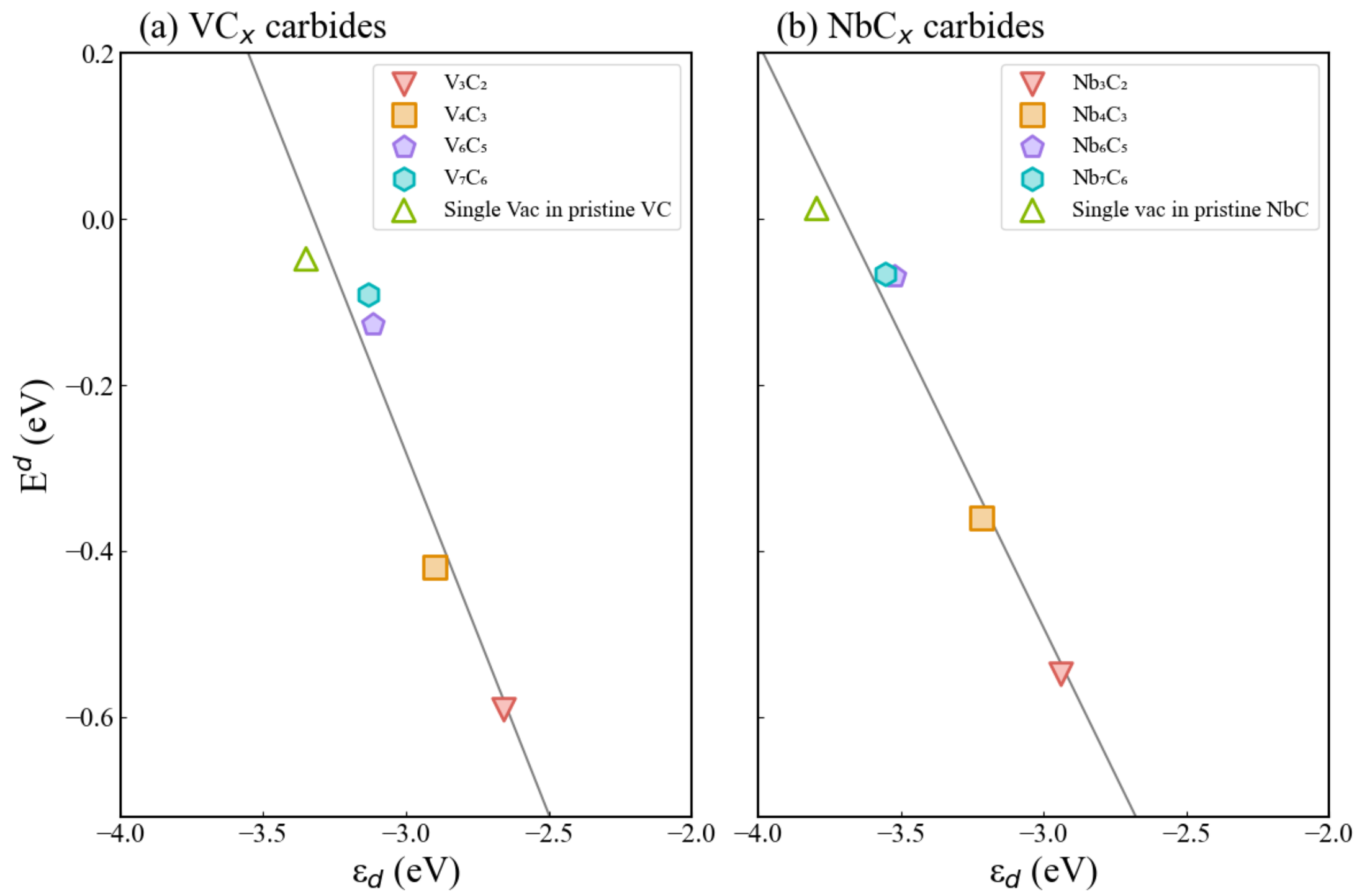

**Figure 6.** The calculated hydrogen dissolution energies ($E^d$) and its correlation with the d band center ($\varepsilon_d$) of (a) $VC_x$ system and (b) $NbC_x$ system

For a quantitative measure of the strength of the hydrogen-metal bonding, pCOHP values have been calculated, with the results shown in **Figure 7**. Upon careful observation of the pCOHP findings for the neighboring V/Nb and hydrogen atoms, hybridization between the hydrogen 1s orbital and the 3*d* (V) and 4*d* (Nb) orbitals are further validated. Notably, a positive pCOHP value denotes bonding contribution, while a negative value signifies antibonding contribution. pCOHP values are typically integrated over an energy range to quantify the bonding strength, resulting in the quantity known as -IpCOHP. This integrated value represents the overall strength of the bond between two atoms. The larger the magnitude of -IpCOHP, the stronger the bonding interaction [62, 63, 88]. In our study, the -IpCOHP values for hydrogen and its adjacent V/Nb atoms were computed. The -IpCOHP value is normalized for each hydrogen insertion by the number of adjacent V/Nb atoms (e.g., hydrogen at a vacancy site is surrounded by six immediate neighboring metal atoms, and thus for the calculation, the integration was performed for six hydrogen-metal bonds before being averaged), providing indication for the strength of a single V/Nb-hydrogen bond. The obtained -IpCOHP values can be found in **Table 2**.

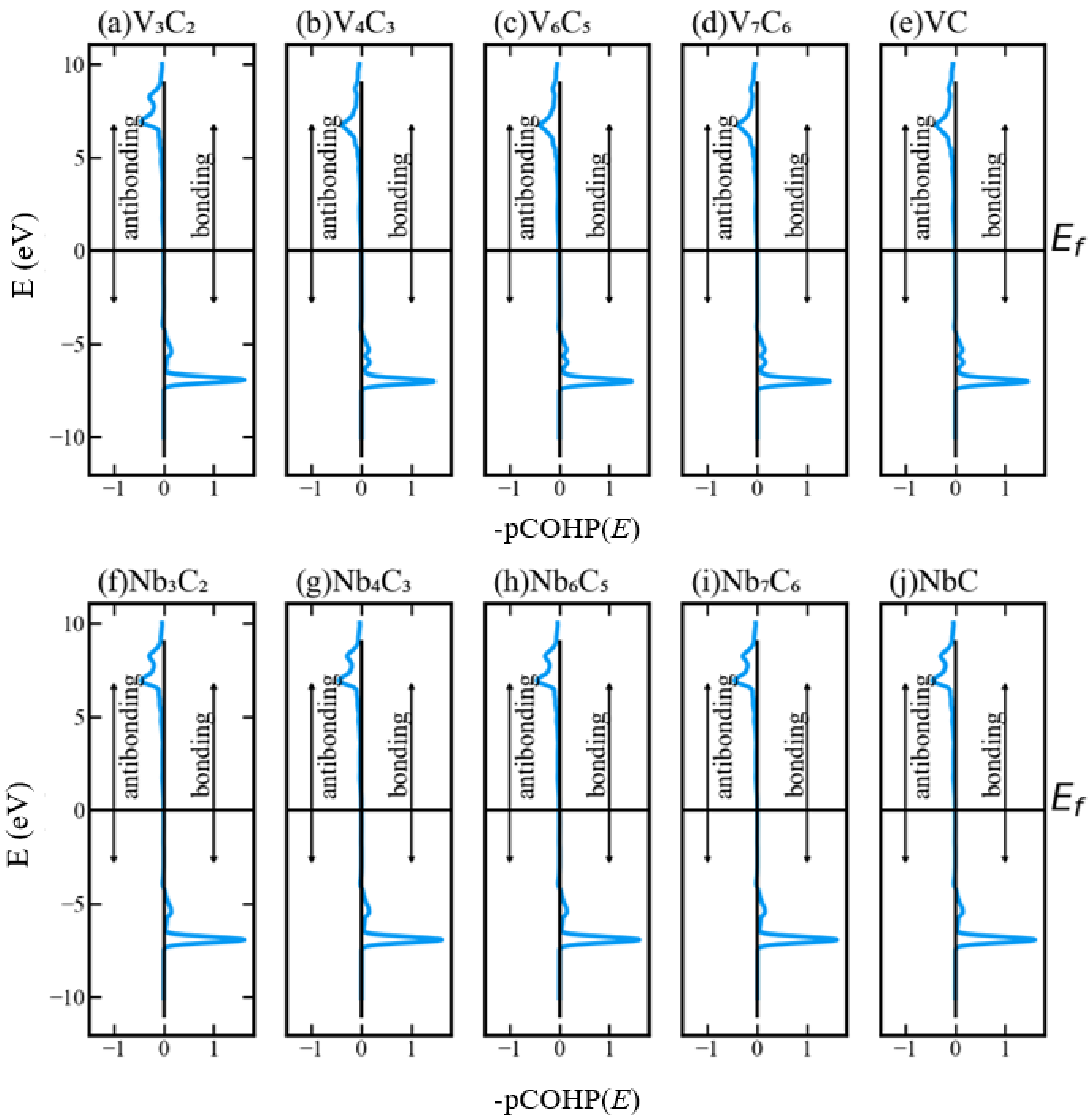

**Figure 7.** The -pCOHP results of bonding between hydrogen and neighboring V/Nb atoms, in different $VC_x$ or $NbC_x$ carbides. Note that (e) and (j) represent hydrogen–V/Nb bonding in a single vacancy.

With the values of $-\mathrm{IpCOHP}$ obtained, it was plotted against the hydrogen dissolution energy, $E^d$, revealing a clear linear relationship as displayed in **Figure 8**, which can be fitted as:

$$E^d = a \times (-\mathrm{IpCOHP}) + c \tag{5}$$

where $a$ and $c$ are fitted parameters displayed in **Table 3**. This linear relationship indicates that hydrogen trapping is primarily driven by bond formation. The enhanced electron density within vacancies prior to hydrogen insertion helps stabilize the carbon atoms and promotes the formation of stronger covalent bonds between hydrogen and the surrounding V/Nb atoms. This mechanism also explains the dependence of hydrogen trapping strength in $VC_x$ and $NbC_x$ on composition, as different carbon content results in different electron density at the vacancy center, which in turn influence bonding behavior.

The above analyses also suggest that the hydrogen adsorption at a $VO_c$ site in $VC_x$ and $NbC_x$ carbides is dominated by the chemical interaction. In this regard, we separately examined the elastic relaxation energy contribution to the hydrogen dissolution energy at a $VO_c$ site and confirmed that this elastic energy contribution is negligible (see **Table 2**).

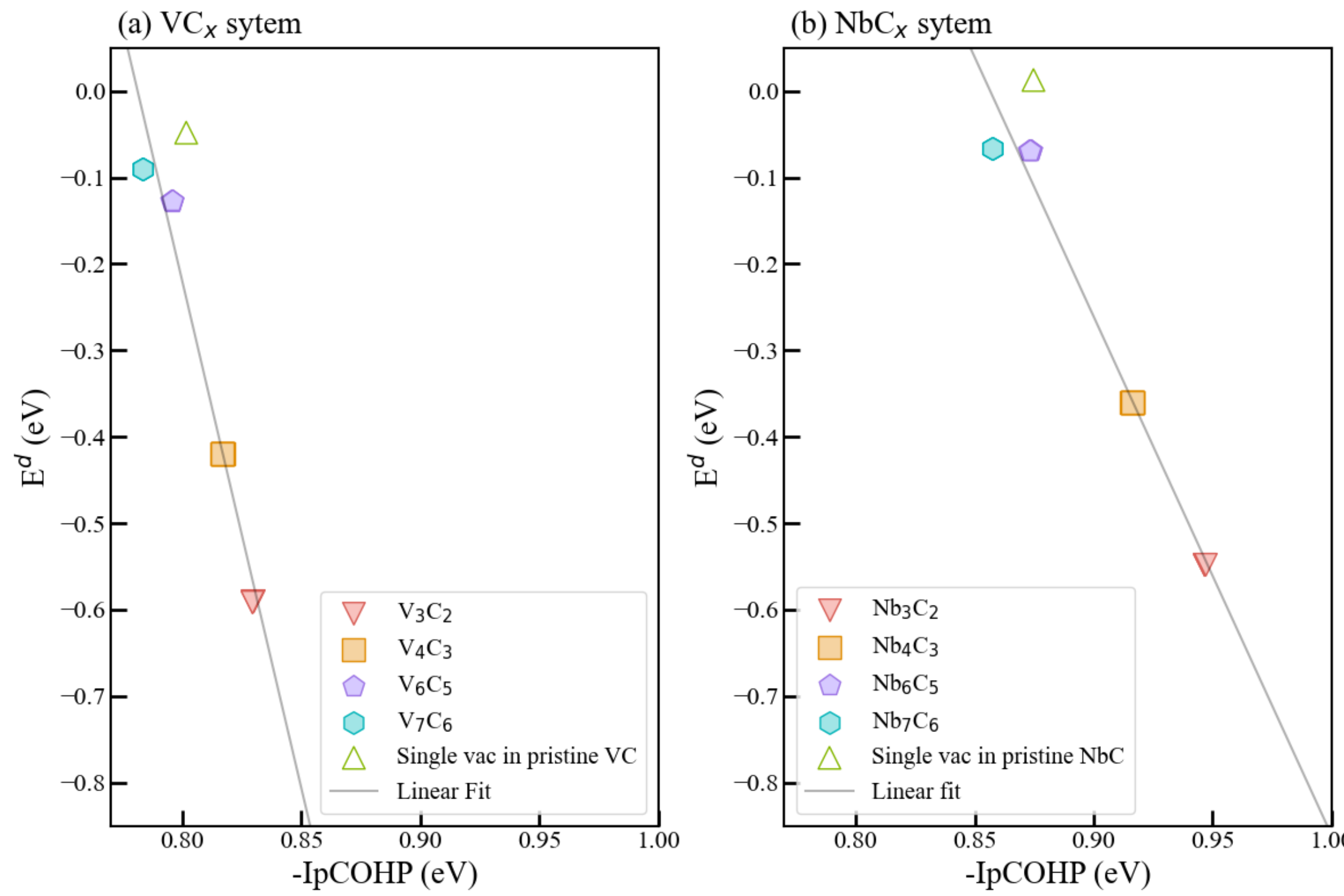

**Figure 8.** Plots of the hydrogen dissolution energy, $E^d$ values at vacancies versus -IpCOHP in (a) $VC_x$ and (b) $NbC_x$ carbides. The green open symbol represents the hydrogen dissolution energy at a single vacancy in the pristine MC (M = V/Nb). The dissolution energies are fitted as $E^d = \mathrm{a} \times (-\mathrm{IpCOHP}) + \mathrm{c}$ with the values of *a* and *c* listed in **Table 3**.

### 3.3 Hydrogen diffusion in sub-stoichiometric carbides

In addition to hydrogen trapping energetics in $VC_x$ or $NbC_x$ carbides, the hydrogen diffusional behaviors and the effect of carbon vacancies were the other important aspect. As discussed early, carbon vacancies in $VC_x$ or $NbC_x$ provide stable trapping sites for hydrogen. Therefore, the possibility of hydrogen migration between those vacancies was examined. Given the high energy states of hydrogen at the normal interstitial sites (other than $VO_c$ sites at vacancy centers) in V/Nb carbides, the ability of hydrogen to migrate between vacancy centers is critical to hydrogen uptake into and escape from the carbides.

**Figure 9** shows the local environments of individual carbon vacancies in different sub-stoichiometric V/Nb carbides. Consistent with our previous work [77], in $M_6C_5$ and $M_7C_6$ (M=V/Nb) carbide compounds, a carbon vacancy forms third-nearest-neighbor (3NN) pairs with its adjacent vacancies along ⟨112⟩ directions (see **Figures 9(c–d)**), while in contrast, carbon vacancies form first-nearest-neighbor (1NN) vacancy pairs along ⟨110⟩ directions in $M_3C_2$ and $M_4C_3$ (**Figures 9(a–b)**). As mentioned earlier, the carbon vacancies in ground-state sub-stoichiometric carbides are intrinsic structural units, and thus the relative spatial locations of neighboring vacancies are well defined, rather than being randomly distributed. Based on the above, hydrogen diffusion, between 3NN vacancies pairs along the ⟨112⟩ direction for $M_7C_6$ and $M_6C_5$ carbides, and between 1NN pairs along the ⟨110⟩ direction for $M_4C_3$ and $M_3C_2$ carbides, was examined.

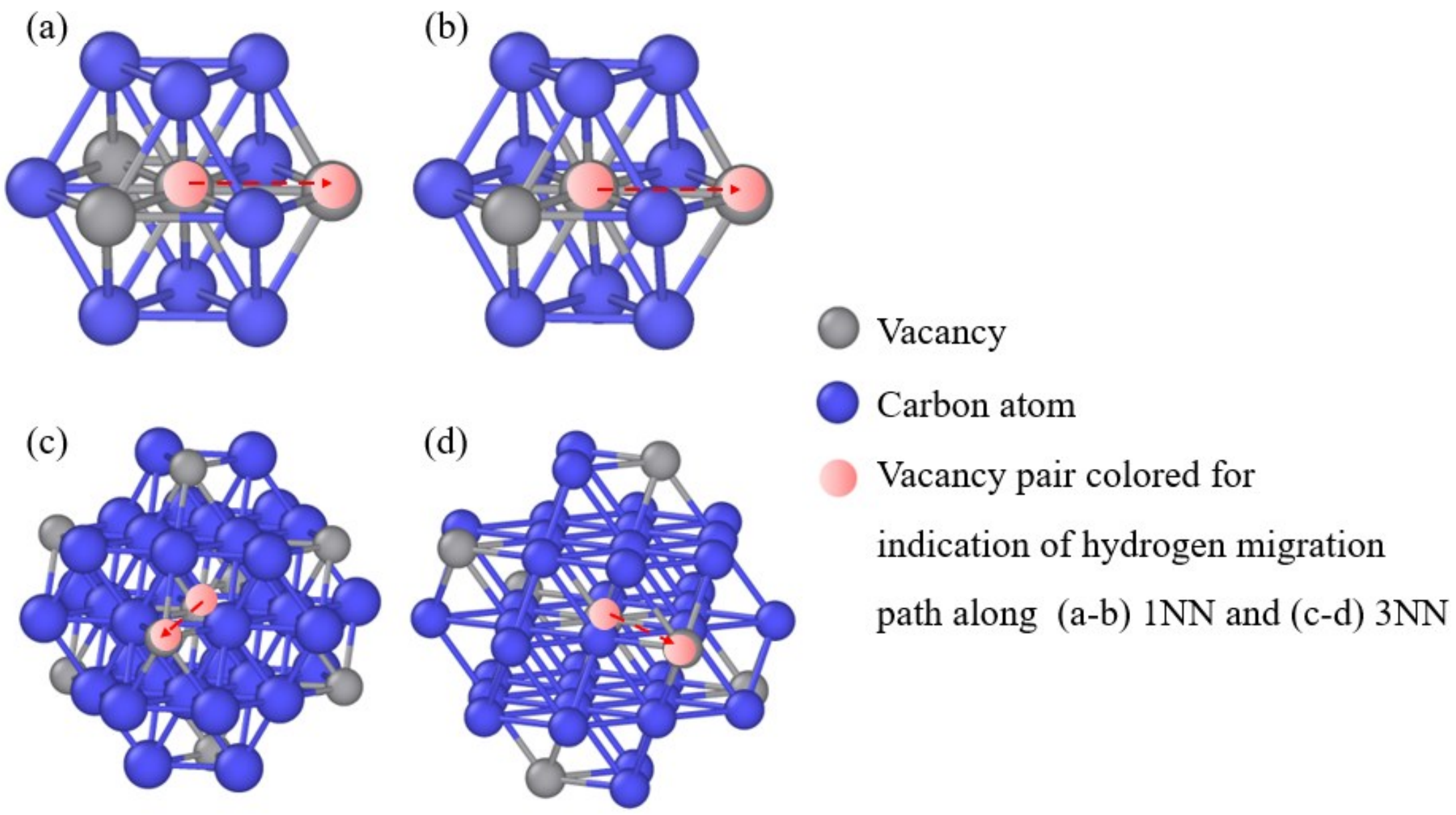

**Figure 9.** This figure shows the arrangement of carbon atoms and carbon vacancies in different $MC_x$ (M=V/Nb) structures. For each central carbon vacancy, we analyzed the distribution of neighboring carbon atoms and vacancies. (a-b) show neighboring 1NN of the central carbon vacancy, including both carbon atoms and vacancies for (a)$M_3C_2$ (b) $M_4C_3$. The central carbon vacancy has carbon vacancies among its 1NN. For $M_6C_5$ and $M_7C_6$, as in Figures (c-d), we analyzed up to 3NN of the central carbon vacancy, and found that the closest vacancy neighbors for the center vacancy are at the 3NN position. Metal atoms are not shown here for clarity.

**Figure 10** presents the relative energy and activation barrier, $E^a$, of hydrogen diffusion along MEPs between vacancies in different sub-stoichiometric $VC_x$/$NbC_x$ carbides. It is apparent that the activation barrier for hydrogen diffusion is dependent on the carbide composition, and more importantly critically affected by the spatial arrangements between neighboring vacancies.

In particular, $E^a$ in **Figure 10(e-f)** shows a significant reduction in the energy barrier after the formation of 1NN vacancy pairs in $M_4C_3$ (0.80 eV for $V_4C_3$ and 0.55eV for $Nb_4C_3$), compared to 3NN pairs in $M_7C_6$ (2.43 eV and 2.33 eV for $V_7C_6$ and $Nb_7C_6$). Detailed values of $E^a$ for the other carbide systems are listed in **Table 2**. The reductions correspond to approximately 67.1% and 76.4% for $V_4C_3$ and $Nb_4C_3$, respectively. This drastic reduction in activation barriers has been suggested to be attributed to the specific vacancy configurations, as well as the energy required for structural dilation along that pathway, as suggested in previous studies [89]. Following these hints from previous studies, we conducted further analysis of the activation barrier, examining the elastic strain and bonding energy contributions at the saddle point. The elastic strain energy contribution, denoted as $E^{\mathrm{ela}}$ (see calculation details in Supplementary Material), was calculated for different carbide systems, with the results shown in **Figure 11(a-b)**. As seen from the figure, the $E^{\mathrm{ela}}$ value is consistently higher in carbide systems containing 3NN pairs than those containing 1NN pairs. Higher $E^{\mathrm{ela}}$ indicates larger deformation as hydrogen passes through the saddle point, thus increasing the activation energy $E^a$. As a result, the elastic deformation renders more resistance to hydrogen diffusion along 3NN paths than along 1NN paths. Meanwhile, the bonding energy contribution at the saddle point can be evaluated using IpCOHP. Here we focus on the interactions between hydrogen and its two nearest V/Nb neighbors, the primary atoms forming bonds with hydrogen. The averaged IpCOHP values are shown in **Figure 11(c–d)**, indicating much stronger hydrogen–metal bonding in 1NN pairs than in 3NN pairs. This enhanced bonding at the saddle point is expected to stabilize hydrogen in the 1NN pathway, thus lowering the energy cost for hydrogen migration. The above results show that the local geometric environment (1NN versus 3NN) modulates both the elastic and bonding components contributing to the activation barrier, providing explanation for the observed variations in $E^a$.

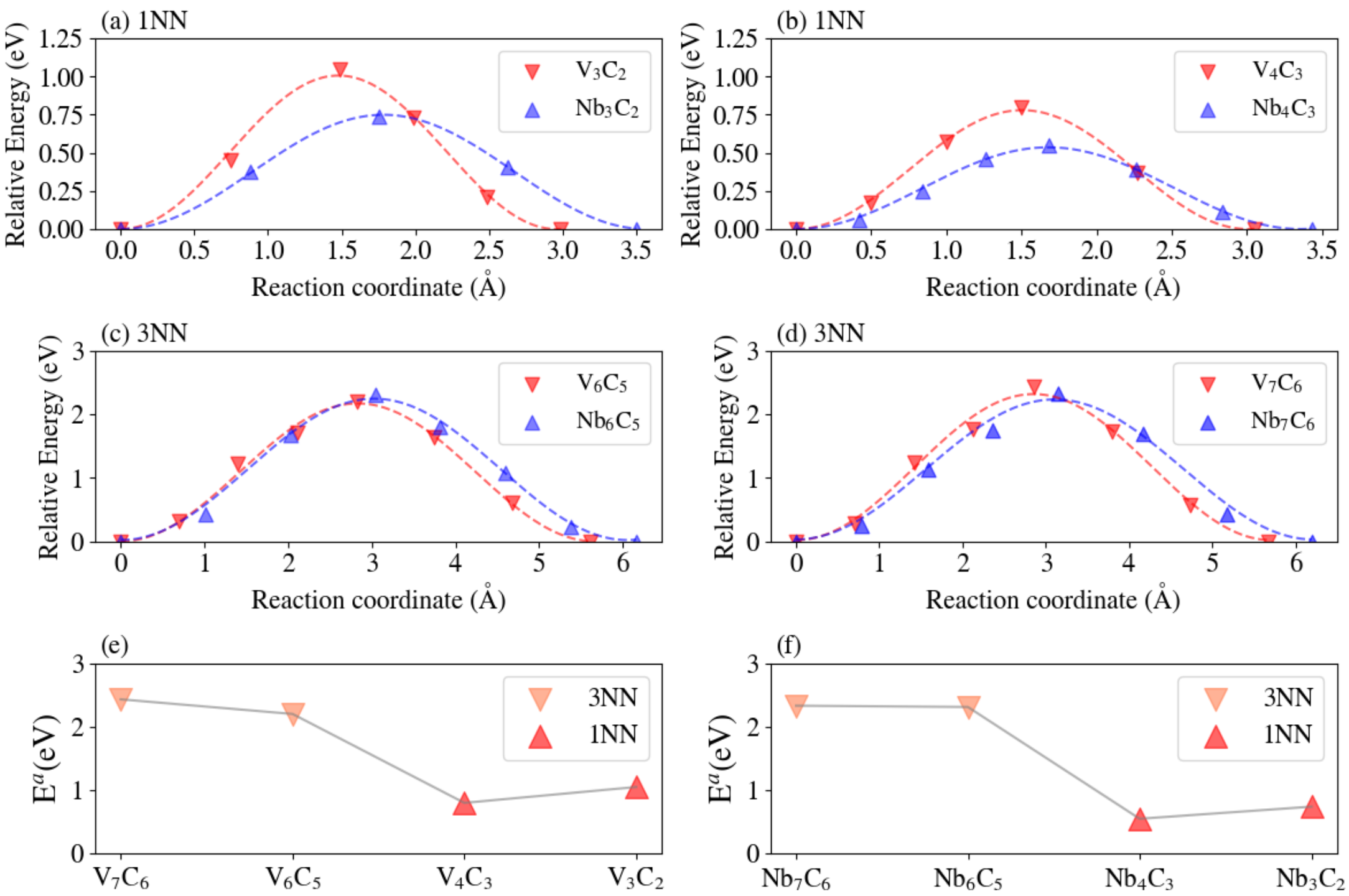

**Figure 10.** NEB results of relative energies during the diffusional process between 1NN in (a) $M_3C_2$ (b) $M_4C_3$ and 3NN in (c)$M_6C_5$, (d) $M_7C_6$. The activation barriers, $E^a$, are displayed for (e) $VC_x$ and (f) $NbC_x$ systems. $E^a$ are 2.43 eV and 2.33 eV for $V_7C_6$ and $Nb_7C_6$ when hydrogen diffuse between 3NN. For $E^a$ between 1NN, the energy barriers are reduced to 0.80 eV (for $V_4C_3$) and 0.55 eV (for $Nb_4C_3$), respectively. The detailed numerical values can be found in **Table 2.**

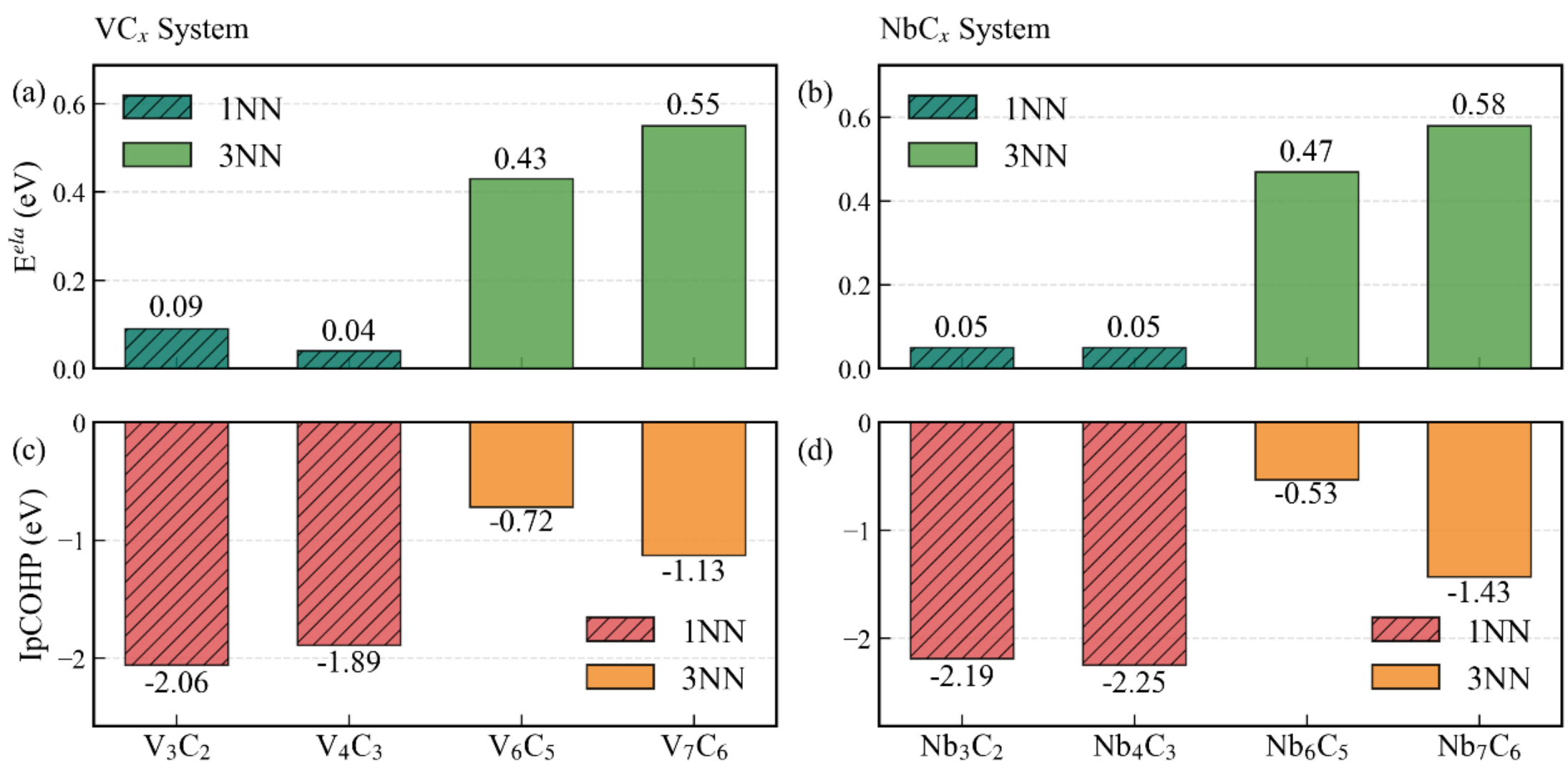

**Figure 11**. Elastic strain energies ($E^{ela}$) induced by hydrogen at the saddle point for the (*a*) $VC_x$ and (*b*) $NbC_x$ carbide systems, together with the analysis of hydrogen-metal bonding at the saddle point based on IpCOHP for the (c) $VC_x$ and (d) $NbC_x$ carbide systems.

Further examining the vacancy configurations in sub-stoichiometric $VC_x/NbC_x$ carbides, it is noted that carbon vacancies are connected via either 1NN or 3NN pairs along ⟨110⟩ and ⟨112⟩ directions respectively on {111} planes. **Figure 12** shows the contour plots of the hydrogen dissolution energy landscapes of the {111} plane for different sub-stoichiometric carbides. From the energy landscapes, the hydrogen diffusion pathways can be inferred. The corresponding relative energy and activation barrier data were collected, also plotted in **Figure 10**. As depicted in the replicated supercell (c.f., **Figure 12**), carbon vacancies align into one-dimensional channels along the $\langle 1\bar{2}\bar{1} \rangle$ directions, being straight and zigzag channels for $M_7C_6$ and $M_6C_5$, and $M_4C_3$ and $M_3C_2$, respectively. Such anisotropic tunnel formation is associated with carbide growth, and is consistent with experimental High-Resolution Transmission Electron

Microscopy (HRTEM) observations of vacancy columns along the same crystallographic direction [90].

We therefore conclude that vacancy interconnections play a key role in hydrogen migration. Variations in vacancy concentration can lead to different pairwise vacancy configurations, which in turn result in distinct hydrogen activation barriers along MEPs. The low hydrogen activation barriers between 1NN vacancy pairs in $M_4C_3$ and $M_3C_2$ (M = V or Nb) carbides, along with the strong trapping energies suggest that they have strong trapping capability yet allowing fast diffusion for hydrogen, making them most effective in HE mitigation.

Our computational findings are also in concert with experimental studies. Takahashi et al. observed significant enhancement in the hydrogen trapping energy as the precipitated VC carbides transform to $V_4C_3$ carbides [32]. Recent APT observations revealed that hydrogen can be stored within the bulk of complex carbides, e.g., (V, Mo)$C_x$ and (Ti, Mo)$C_x$, when considerable concentration of vacancies is present [49, 50]. A comparative study between $TiC_x$ and (Ti, Mo)$C_x$ showed that Mo alloying results in a smaller $x$ value and thus higher vacancy concentration compared to $TiC_x$, accompanying which hydrogen trapping changes from being predominantly at the carbide/matrix interface to being primarily within the carbide interior [49]. Such transition aligns well with our computational results that increasing vacancy concentration not only enhance hydrogen trapping strength but also facilitate hydrogen entry into carbides. Our computational work compliments experimental studies by providing precise determination of carbon vacancy structures and quantification of vacancy content, addressing the inherent challenges in experiments as noted by Liu et al. [49].

On the other hand, it is important to note the limitation of the present study, which only focuses on hydrogen trapping and diffusion in the carbide interior. The effectiveness of carbides in mitigating HE also depends on the thermodynamics and kinetics of hydrogen at the carbide/steel interface, which not only plays a critical role in affecting hydrogen trapping but also prescribes the accessibility of carbide interior to hydrogen. Although it is evident from experiments that hydrogen may reside inside the sub-stoichiometric transition-metal carbides [32, 49, 50], there also exist many studies suggesting carbide/steel interfaces as the preferred locations for hydrogen trapping [39, 91, 92]. For instance, Shi et al. demonstrated strong hydrogen trapping of hydrogen at the semi-coherent Nb carbide/$\alpha$-iron interface, characterized by a substantial dissolution energy of -0.80 eV [91]. Meanwhile, accompanying the potential for trapping hydrogen, those interfaces can certainly exert significant influence on hydrogen diffusion. In fact, it can be envisioned that the relevance of carbides in mitigating HE would be highly sensitive to the barriers of hydrogen diffusion at those interfaces as a low or high barrier at the interface would effectively facilitate or inhibit hydrogen exchange between carbides and steel matrix.

Nevertheless, the competition between hydrogen trapping at the carbide/steel interface and carbide interior is far from being well understood. This knowledge deficit is partially attributed to the structural complexity at carbide/steel interfaces (e.g., interface orientation, coherency and existence of other defects), which is a topic of paramount importance warranting dedicated future studies beyond our present work.

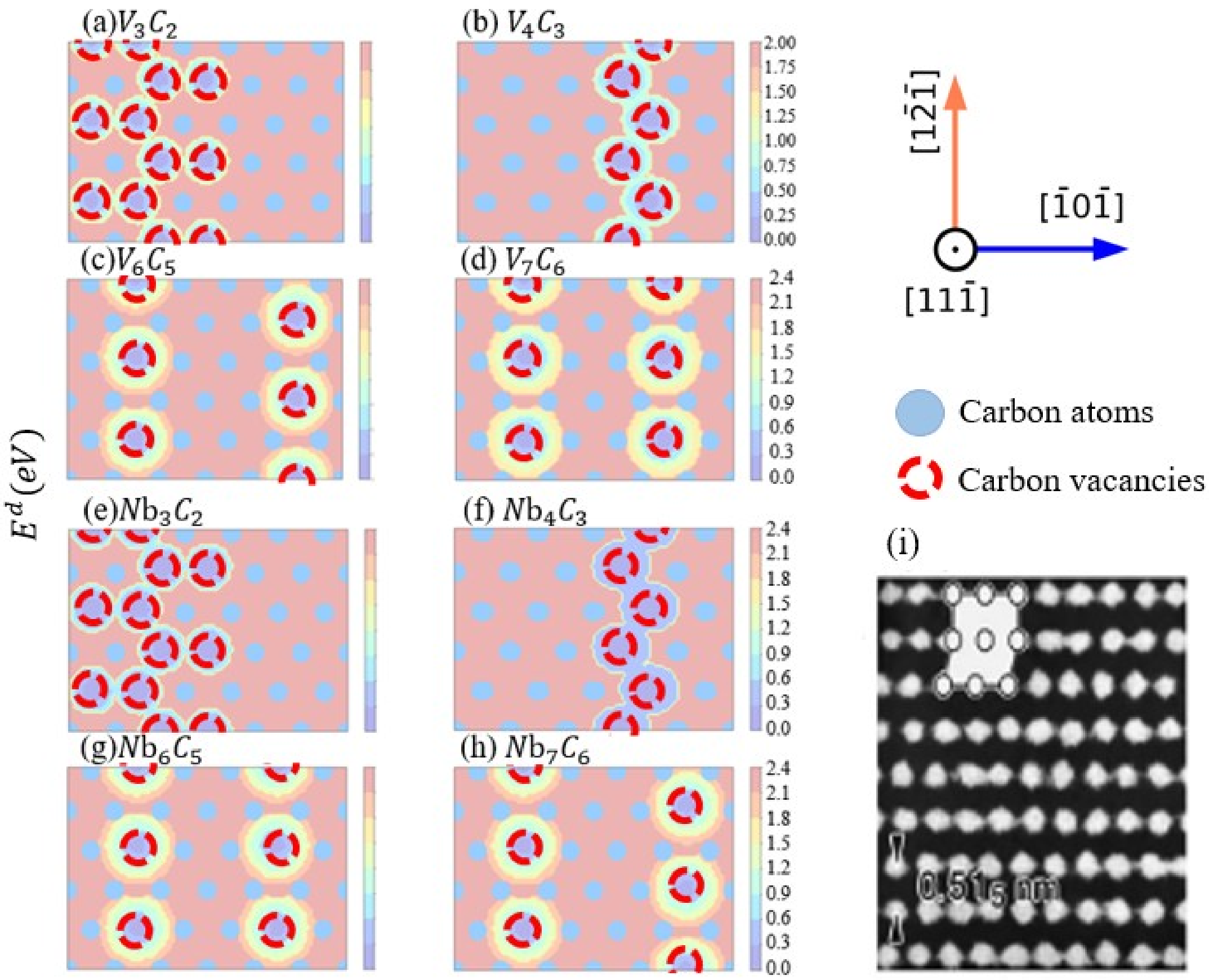

**Figure 12**. Contour plots showing the hydrogen dissolution energy landscapes on {111} planes for (a) $V_3C_2$, (b) $V_4C_3$, (c) $V_6C_5$, (d) $V_7C_6$, (e) $Nb_3C_2$, (f) $Nb_4C_3$, (g) $Nb_6C_5$, (h) $Nb_7C_6$. Here, blue circles represent carbon atoms; metal atoms are not shown here for clarity. (i) HRTEM image of vacancy columns viewed along the $\langle 1\bar{2}\bar{1} \rangle$ direction. Reprinted with permission from Ref. [90]

## 4. Conclusion

In summary, employing systematic first-principles calculations, hydrogen trapping and diffusion characteristics in sub-stoichiometric $VC_x$/$NbC_x$ carbides were investigated. Below the main findings are listed.

1. The stable trapping sites for hydrogen in sub-stoichiometric $VC_x$/$NbC_x$ carbides were found to coincide with the carbon vacancy centers, where hydrogen exhibits a lower dissolution energy than that for hydrogen at an interstitial site in pristine bcc Fe or at a single vacancy in pristine VC/NbC carbides.
2. The hydrogen dissolution energy in sub-stoichiometric carbides was also shown to strongly composition dependent, decreasing with increasing carbon vacancy content, with the hydrogen trapping strength transition from weak (reversible) trapping to strong (irreversible) trapping.
3. The hydrogen dissolution energy (trapping strength) was shown to be largely prescribed by the *d*-band center of the metal atoms in different carbide systems before hydrogen insertion. As the carbon vacancy content increases, the *d*-state shifts upward, resulting in stronger hydrogen–metal interactions and therefore stronger trapping. These trends are consistent with prior theoretical and experimental observations on transition-metal carbides, underscoring the critical role of local electronic structure in determining hydrogen stability.
4. Hydrogen diffusion characteristics in sub-stoichiometric $VC_x$/$NbC_x$ carbides also show strong composition dependence. The energy barrier for hydrogen diffusion between neighboring trap sites was shown to decrease significantly as the arrangement of carbon vacancies within carbides change from the third-nearest-neighbor (3NN) form to the first-nearest-neighbor (1NN) form.
5. The 1NN vacancy pattern was demonstrated to enable interconnected low-barrier hydrogen diffusion channels in the carbides. This result, along with the trapping energy analysis suggest $M_4C_3$ and $M_3C_2$ (M = V or Nb), of strong trapping capability yet allowing fast diffusion for hydrogen, as being most effective in HE mitigation.

The present study contributes novel insights into hydrogen trapping and diffusion within sub-stoichiometric V/Nb carbides. The knowledge gained is expected to contribute valuable guidance for process and composition innovation of high-strength alloy steels for better HE resistance. In addition, the importance and necessity of future studies to better understand hydrogen trapping and diffusion at carbide/steel matrix interfaces are also acknowledged.

## Acknowledgements

The authors sincerely thank the Natural Sciences and Engineering Research Council of Canada (NSERC) Alliance and Discovery Grants program for generously funding this research. We also extend our heartfelt gratitude to Digital Alliance for providing invaluable computational resources, which were essential for the simulations and data analysis conducted in this study. Special thanks are given to Dr. Jie Hou for his insightful suggestions that greatly contributed to this work.